\documentclass[superscriptaddress,onecolumn,english,floatfix,aps,pra,amsmath,amssymb,longbibliography]{revtex4-1}
\usepackage[T1]{fontenc}
\usepackage[utf8]{inputenc}
\usepackage{babel}
\usepackage{multirow}
\usepackage{comment}
\usepackage{color}
\usepackage{array}
\usepackage{enumitem}
\usepackage{times}
\usepackage[table,xcdraw]{xcolor}
\usepackage{epsfig}
\usepackage{subfigure}
\usepackage{amsmath}
\usepackage{amssymb}
\usepackage{natbib}
\usepackage{graphicx}
\usepackage{natbib}
\DeclareMathOperator{\sech}{sech}

\usepackage[colorlinks=true]{hyperref}
\hypersetup{citecolor=blue,linkcolor=blue,urlcolor=blue}

\begin{document}
	
\title{Squeezing equivalence of quantum harmonic oscillators under different frequency modulations}
	
\author{Stanley S. Coelho}
\email[Corresponding Author: ]{stanley.coelho@icen.ufpa.br}
\affiliation{Faculdade de F\'{i}sica, Universidade Federal do Par\'{a}, 66075-110, Bel\'{e}m, Par\'{a}, Brazil}	
\author{Lucas Queiroz}
\email{lucas.queiroz@ifpa.edu.br}
\affiliation{Faculdade de F\'{i}sica, Universidade Federal do Par\'{a}, 66075-110, Bel\'{e}m, Par\'{a}, Brazil}
\affiliation{Instituto Federal de Educa\c{c}\~{a}o, Ci\^{e}ncia e Tecnologia do Par\'{a}, Campus Rural de Marab\'{a}, 68508-970, Marab\'{a}, Par\'{a}, Brazil}
\author{Danilo T. Alves}
\email{danilo@ufpa.br}
\affiliation{Faculdade de F\'{i}sica, Universidade Federal do Par\'{a}, 66075-110, Bel\'{e}m, Par\'{a}, Brazil}

\date{\today}


\begin{abstract}

The papers by Janszky and Adam [Phys. Rev. A {\bf 46}, 6091 (1992)] and Chen \textit{et al.} [Phys. Rev. Lett. {\bf 104}, 063002 (2010)] are examples of works where one can find the following equivalences: 
quantum 
harmonic oscillators subjected to different time-dependent frequency modulations, during a certain time interval $\tau$, exhibit exactly the same final null squeezing parameter ($r_f=0$).
In the present paper, we discuss a more general case of squeezing equivalence, where the final squeezing parameter can be non-null ($r_f\geq0$). 
We show that when the interest is in controlling the forms of the frequency modulations, but keeping free the choice of the values of $r_f$ and $\tau$, this in general demands numerical calculations to find these values leading to squeezing equivalences (a particular case of this procedure recovers the equivalence found by Jansky and Adams).
On the other hand, when the interest is not in previously controlling the form of these frequencies, but rather $r_f$ and $\tau$ (and also some constraints, such as minimization of energy), one can have analytical solutions for these frequencies leading to squeezing equivalences (particular cases of this procedure are usually applied in problems of shortcuts to adiabaticity, as done by Chen \textit{et al.}). 
In this way, this more general squeezing equivalence discussed here is connected to recent and important topics in the literature as, for instance, generation of squeezed states and the obtaining of shortcuts to adiabaticity. 

\end{abstract}
\maketitle


\section{Introduction}\label{sec:introduction}

The time-dependent quantum harmonic oscillator (TDHO) is relevant in modeling several problems in physics \cite{Husimi-PTP-1953-II, Lewis-JMP-1969, Pedrosa-PRA-1997,Pedrosa-PRA-1997-singular-perturbation, Ciftja-JPA-1999}, and has been investigated in different situations, such as in the description of the interaction between a spinless charged quantum particle and a time-dependent external classical electromagnetic field \cite{Dodonov-PLA-1994, Aguiar-JMP-2016, Dodonov-JRLR-2018}, in the quantum particle motion in traps \cite{Brown-PRL-1991,Alsing-PRL-2005, Menicucci-PRA-2007,Mihalcea-PS-2009, Mihalcea-AP-2022, Mihalcea-PR-2023}, as well as in the quantization of the free electromagnetic field in nonstationary media \cite{Pedrosa-PRL-2009, Choi-PRA-2010, Pedrosa-PRA-2011, Unal-AP-2012, Lakehal-SR-2016}. 
In quantum circuit electrodynamics, a double superconducting quantum interference device can be modeled as a TDHO with time-dependent frequency, consisting of an important system for the study of the dynamical Casimir effect \cite{Fujii-PRB-2011,Wilson-Nature-2011,Lahteenmaki-PNAS-2013}.
In the description of scalar fields in Friedmann-Robertson-Walker spacetime and in the analysis of particle production in de Sitter spacetime, TDHOs have also been considered, since the Hamiltonian of the field modes can be mapped onto the Hamiltonian of a TDHO with time-dependent mass and frequency \cite{Pedrosa-PRD-2004, Pedrosa-PLB-2007, Lopes-JMP-2009, Greenwood-IJMP-2015}.
In the context of shortcuts to adiabaticity, TDHOs are also utilized to describe the fast frictionless cooling of atoms and quantum gases in harmonic traps \cite{Salamon-PCCP-2009, Chen-PRL-2010,Chen-PRA-2010,Stefanatos-PRA-2010, Choi-PRA-2012,Choi-PRA-2013,Odelin-RMP-2019,Beau-Entropy-2020, Huang-Chaos-2020,Dupays-PRR-2021}.

An interesting feature of TDHOs is that they can be used to create squeezed states. 
The main characteristic of these states is the reduction in the variance of a given quadrature, implying an increase in the variance of the corresponding conjugate quadrature (for example, position and momentum) \cite{Guerry-Quantum-Optics-2005}.
This makes possible to reduce the noise of a given measurement beyond the standard quantum limit (also known as the Heisenberg limit), leading to important applications in the area of quantum sensing, such as the improvement on the detection of gravitational waves \cite{Degen-RMP-2017,Pezze-RMP-2018}.
A particular case of a TDHO is one that presents sudden frequency jumps \cite{Janszky-OC-1986,Graham-JMO-1987,Yi-PRA-1988,Janszky-PRA-1989,Lo-JPA-1990,Agarwal-PRL-1991,Janszky-PRA-1992,Averbukh-PRA-1994,Kiss-PRA-1994,Feng-CTP-1997,Cessa-PLA-2003,Ahmadi-JPB-2006,Galve-PRA-2011,Hoffmann-PRE-2013,Matsuo-PB-2015,Alonso-Nature-2016,Rashid-PRL-2016,Joshi-SR-2017,Tibaduiza-BJP-2020,Xin-PRL-2021,Cosco-PRA-2021,Coelho-Entropy-2022}.
In Ref. \cite{Janszky-OC-1986}, Janszky and Yushin showed that these jumps create squeezed states (see also Refs. \cite{Graham-JMO-1987,Yi-PRA-1988,Janszky-PRA-1989,Lo-JPA-1990,Agarwal-PRL-1991,Janszky-PRA-1992,Averbukh-PRA-1994,Kiss-PRA-1994,Feng-CTP-1997,Cessa-PLA-2003,Ahmadi-JPB-2006}). 
While this procedure is still widely used \cite{Galve-PRA-2011,Hoffmann-PRE-2013,Matsuo-PB-2015,Alonso-Nature-2016,Rashid-PRL-2016,Joshi-SR-2017,Tibaduiza-BJP-2020,Xin-PRL-2021,Cosco-PRA-2021,Coelho-Entropy-2022}, it is worth noting that sudden jumps in frequency are not the only way to create squeezed states in TDHOs. For example, in Refs. \cite{Ma-PRA-1989, Aliaga-PRA-1990, Mang-CTP-1997,Galve-PRA-2009,Tibaduiza-PS-2020,Tibaduiza-JPB-2021}, it is shown that squeezing will be generated even for smoothly varying time-dependent frequencies.

One can cite Refs. \cite{Chen-PRL-2010,Janszky-PRA-1992} as examples of works where we can find the following cases of equivalence: TDHOs with the same initial frequency $\omega_0$, the same final frequency $\omega_f$, and the same null initial squeezing parameter ($r_{0}=0$), subjected to different intermediary frequency modulations during a certain time interval $\tau$, exhibit exactly the same final null squeezing parameter ($r_f=0$).
More specifically, in Ref. \cite{Janszky-PRA-1992}, Janszky and Adam found an interesting equivalence: a TDHO, under a sequence of two sudden frequency jumps, from $\omega_0$ (for $t\leq0$) to $\omega_1$, and returning to $\omega_0$ (for $t>\tau$), exhibits, for $t>\tau=s\pi/\omega_1$ ($s\in\mathbb{N}$), a null squeezing parameter ($r_f=0$), exactly as occurs for a harmonic oscillator whose frequency would remain constant. 
In Ref. \cite{Chen-PRL-2010}, Chen \textit{et al.} have studied ways of accelerating the frictionless cooling process of atoms in harmonic traps based on a single parameter: the angular frequency $\omega(t)$ of the TDHO.
This procedure is related to problems of shortcuts to adiabaticity, as the system is prepared to maintain the same initial and final states (less than a global phase factor) and energies, as an adiabatic process, but in a shorter time than that needed to reproduce a conventional adiabatic process \cite{Chen-PRL-2010}.
Thus, since the initial and final populations are the same of an usual adiabatic process \cite{Chen-PRL-2010}, and the adiabatic processes are related to processes with null squeezing \cite{Choi-PRA-2012,Choi-PRA-2013,Beau-Entropy-2020,Galve-PRA-2009}, then different protocols that seek shortcuts to adiabaticity are also equivalent in the sense that the final squeezing parameters are null. 

In the present paper, we discuss a more general case of squeezing equivalence, where the final squeezing parameter can be non-null ($r_f\geq0$), and valid for discontinuous or continuous frequency modulations (as far as we know, this more general case of squeezing equivalence has never been discussed in the literature, and, in addition, the nomenclature ``squeezing equivalence'' itself is being proposed in the present article).
Given that the quantum fluctuations of the position and momentum operators, quantum fluctuations of the energy value, mean number of excitations and its fluctuations, and the transition probabilities between different states, depend on the squeezing parameter, the more general case of squeezing equivalence presented here implies that all these quantities will also be the same for the TDHOs, for $t>\tau$.
Taking as basis the Lewis-Riesenfeld (LR) dynamical invariant method \cite{Lewis-JMP-1969}, we show that when the interest is in controlling the forms of intermediary frequency modulations, but keeping free $r_f$ and $\tau$, this in general leads to numerical solutions to find values of $r_f$ and $\tau$, which lead to squeezing equivalence (a particular case of this procedure recovers the equivalence found by Jansky and Adams \cite{Janszky-PRA-1992}).
On the other hand, when the interest is not in previously controlling the form of the intermediary frequencies, but rather $r_f$ and $\tau$ (and also some constraints, such as frequency limitations and minimization of energy), one can have analytical solutions to find these frequencies leading to squeezing equivalence (particular cases of this procedure are usually applied in problems of shortcuts to adiabaticity, as done by Chen \textit{et al.} \cite{Chen-PRL-2010}). 

The paper is organized as follows. 
In Sec. \ref{sec:method}, we make a brief review of the solution of the Schrödinger equation associated with the TDHO with time-dependent frequency, obtained via the LR method \cite{Lewis-JMP-1969,Pedrosa-PRA-1997,Pedrosa-PRA-1997-singular-perturbation}. 
We also introduce expressions for the squeezing parameter and squeezing phase, quantum fluctuations of the position, momentum, energy and number of excitations, and the transition probabilities between different states. 
In Sec. \ref{sec:model}, we investigate the dynamics of a TDHO with initial frequency $\omega_{0}$, which evolves to a time-dependent intermediary frequency $\omega_{\text{int}}(t)$ and, after a certain time interval $\tau$, assumes the final frequency $\omega_{f}$.
In Sec. \ref{sec:equivalence}, we obtain the conditions that allow us to configure a set of TDHOs, subject to different intermediary frequency modulations, so that they result in a same value for $r_f$ ($r_f\geq0$) and, consequently, in the same physical quantities that depend on it, after the instant $\tau$.
In Sec. \ref{sec:aplicações}, we illustrate applications of this squeezing equivalence.
In Sec. \ref{sec:final}, we present our final remarks.


\section{Analytical Method}\label{sec:method}

In this section, by means of the LR method \cite{Lewis-JMP-1969}, we make a brief review of the solution of the Schrödinger equation associated with the TDHO with time-dependent frequency, introducing the squeezing parameter and quantities related to it. These topics are well known and
found in the literature (see, for instance, Refs. \cite{Janszky-OC-1986,Agarwal-PRL-1991,Yi-PRA-1988,Pedrosa-PRA-1997,Pedrosa-PRA-1997-singular-perturbation,Tibaduiza-PS-2020}), being the novelty of the present paper in the introduction and
discussion of a more general case of squeezing equivalence, which allows different TDHOs to be configured in such a way that they have the same non-null final squeezing parameter.

\subsection{The wave function of the time-dependent quantum harmonic oscillator}\label{sec:wave function}


Let us consider the Schrödinger equation 
\begin{eqnarray}
	\label{eq:equação de Schrodinger}
	i\hbar\frac{\partial\Psi(x,t)}{\partial t}=\hat{H}(t)\Psi(x,t),
\end{eqnarray}
for a one-dimensional TDHO with mass $m_0$ and time-dependent angular frequency $ \omega(t) $, whose Hamiltonian operator $\hat{H}(t)$ is
\begin{eqnarray}
	\label{eq:Hamiltoniano}
	\hat{H}(t)=\frac{\hat{p}^{2}}{2m_{0}}+\frac{1}{2}m_{0}\omega(t)^{2}\hat{x}^{2},
\end{eqnarray}
where $\hat{x}$ and $\hat{p}$ are the position and momentum operators, respectively.
According to the LR method \cite{Lewis-JMP-1969,Pedrosa-PRA-1997,Pedrosa-PRA-1997-singular-perturbation}, the general solution of Eq. \eqref{eq:equação de Schrodinger} is a superposition of orthonormal “expanding modes”
\begin{eqnarray}
	\Psi(x,t)=\sum_{n=0}^{\infty}C_{n}\Psi_{n}(x,t),
	\label{eq:Psi}
\end{eqnarray}
in which the time-independent coefficients $ C_n $ depend only on the initial conditions and $\Psi_{n}(x,t)$ is given by
\begin{eqnarray}
	\Psi_{n}(x,t)=\exp[i\alpha_{n}(t)]\Phi_{n}(x,t),
\end{eqnarray}
where $\alpha_{n}(t)$ are phase functions, defined by 
\begin{eqnarray}
	\alpha_{n}(t)=-\left(n+\frac{1}{2}\right)\int_{0}^{t}\frac{dt^{\prime}}{m_{0}\rho(t^{\prime})^{2}},
	\label{eq:alpha_{n}}
\end{eqnarray}
and the functions $\Phi_{n}(x,t)$ are given by
\begin{eqnarray}
\Phi_{n}(x,t)=\frac{1}{\sqrt{2^{n}n!}}\biggl[\frac{1}{\pi\hbar\rho(t)^{2}}\biggr]^{\frac{1}{4}}\exp\biggl\{\frac{im_{0}}{2\hbar}\biggl[\frac{\dot{\rho}(t)}{\rho(t)}+\frac{i}{m_{0}\rho(t)^{2}}\biggr]x^{2}\biggr\}{\cal H}_{n}\biggl[\frac{x}{\hbar^{\frac{1}{2}}\rho(t)}\biggr],
\label{eq:phi_{n}}
\end{eqnarray}
with ${\cal H}_{n}$ being the Hermite polynomials of order $ n $. 
The function $ \rho(t) $ in Eqs. \eqref{eq:alpha_{n}} and \eqref{eq:phi_{n}} is a real parameter which is solution of the Ermakov-Pinney equation \cite{Pinney-PAMS-1950,Carinena-IJGMMP-2009,Lima-JMO-2009,Mancas-AMC-2015}:
\begin{eqnarray}
	\label{eq:equação de Ermakov-Pinney}
	\ddot{\rho}(t)+\omega(t)^{2}\rho(t)=\frac{1}{m_{0}^{2}\rho(t)^{3}}.
\end{eqnarray}
For the case in which the frequency is always constant [$\omega(t)=\omega_{0}$], the solution of Eq. \eqref{eq:equação de Ermakov-Pinney} is $\rho(t)=\rho_0$, in which
\begin{eqnarray}
	\label{eq:rho0}
	\rho_0=\frac{1}{\sqrt{m_{0}\omega_{0}}},
\end{eqnarray}
so that $\Psi_{n}(x,t)$ falls back to the wave function $\Psi_{n}^{(0)}(x,t)$ of a quantum harmonic oscillator with time-independent mass and frequency \cite{Sakurai-Quantum-Mechanics-2021}. 
An enlightening discussion of the connection between the Ermakov-Pinney equation \eqref{eq:equação de Ermakov-Pinney} and classical mechanics is found in Refs. \cite{Huang-Chaos-2020,Anderson-JO-1994}.


\subsection{Squeezing parameter and related quantities}\label{sec:squeezing parameters}

It is well known that the quantum states $ \Psi_{n}(x,t) $ of the TDHO are squeezed (see, for instance, Refs. \cite{Pedrosa-PRA-1997,Pedrosa-PRA-1997-singular-perturbation}). 
Thus, we can define the squeezing parameter $r(t)$ and the squeezing phase $\phi(t)$ [with $r(t)\geq 0 $ and $ 0\leq\phi(t)\leq 2\pi$] in terms of the parameter $\rho(t)$, so that they can be written in the instantaneous basis \cite{Tibaduiza-JPB-2021} as 
\begin{eqnarray}
	r(t)=\cosh^{-1}\{[\lambda(t)]^{\frac{1}{2}}\},
	\label{eq:r}
\end{eqnarray}
where
\begin{eqnarray}
	\lambda(t)=\frac{m_{0}^{2}\dot{\rho}(t)^{2}+\frac{1}{\rho(t)^{2}}+m_{0}^{2}\omega(t)^{2}\rho(t)^{2}+2m_{0}\omega(t)}{4m_{0}\omega(t)},
	\label{eq:lambda(t)}
\end{eqnarray}
and
\begin{eqnarray}
	\phi(t)=\cos^{-1}\left\{ \frac{1+m_{0}\omega(t)\rho(t)^{2}-2\cosh^{2}[r(t)]}{2\sinh[r(t)]\cosh[r(t)]}\right\}.
	\label{eq:phi}
\end{eqnarray}

Defining the variance of a given operator $\hat{{\cal O}}$ in the state $\Psi_{n}(x,t)$ as $\sigma_{\hat{{\cal O}}}^{2}(n,t)=[\langle\hat{{\cal O}}^{2}\rangle(n,t)]-[\langle\hat{{\cal O}}\rangle(n,t)]^{2}$, in which
\begin{eqnarray}
	\langle\hat{{\cal O}}\rangle(n,t)=\int_{-\infty}^{+\infty}dx\Psi_{n}^{*}(x,t)\hat{{\cal O}}\Psi_{n}(x,t),
\end{eqnarray}
we can write the variances of the position [$\sigma_{\hat{x}}^{2}(n,t)$] and momentum [$\sigma_{\hat{p}}^{2}(n,t)$] operators in terms of $r(t)$ and $\phi(t)$ as 
\begin{align}
	\sigma_{\hat{x}}^{2}(n,t)\;= & \;\,\{\sinh^{2}[r(t)]+\cosh^{2}[r(t)]+2\sinh[r(t)]\cosh[r(t)]\cos[r(t)]\}\overline{\sigma}_{\hat{x}}^{2}(n,t),
	\label{eq:<x^{2}>}\\
	\sigma_{\hat{p}}^{2}(n,t)\;= & \;\,\{\sinh^{2}[r(t)]+\cosh^{2}[r(t)]-2\sinh[r(t)]\cosh[r(t)]\cos[r(t)]\}\overline{\sigma}_{\hat{p}}^{2}(n,t),
	\label{eq:<p^{2}>}
\end{align}
where $\overline{\sigma}_{\hat{x}}^{2}(n,t)$ and $\overline{\sigma}_{\hat{p}}^{2}(n,t)$ are defined respectively by 
\begin{align}
	\overline{\sigma}_{\hat{x}}^{2}(n,t)=\biggl(n+\frac{1}{2}\biggr)\frac{\hbar}{m_{0}\omega(t)},\\
	\overline{\sigma}_{\hat{p}}^{2}(n,t)=\biggl(n+\frac{1}{2}\biggr)\hbar m_{0}\omega(t).
\end{align}
Note that $\langle\hat{x}\rangle(n,t)=\langle\hat{p}\rangle(n,t)=0$, just as in the case where $\omega(t)=\omega_{0}$ for any time interval.
In addition, we can also define the mean energy $E(n,t)=\langle\hat{H}(t)\rangle(n,t)$: 
\begin{eqnarray}
	E(n,t)=\cosh[2r(t)]\overline{E}(n,t),
	\label{eq:<H>}
\end{eqnarray}
where $\overline{E}(n,t)=(n+1/2)\hbar\omega(t)$, and the energy variance $\sigma_{\hat{H}}^{2}(n,t)$ as \cite{Beau-Entropy-2020,Galve-PRA-2009}  
\begin{eqnarray}
	\sigma_{\hat{H}}^{2}(n,t)=\frac{\hbar^{2}\omega(t)^{2}}{2}(n^{2}+n+1)\sinh^{2}[2r(t)].
	\label{eq:var-E}
\end{eqnarray}
One can also determine the mean number of excitations $N(n,t)$, and its variance $\sigma_{\hat{N}}^{2}(n,t)$, for a system subjected to this time-varying potential. These quantities are given respectively by \cite{Kim-PRA-1989,Marian-PRA-1992,Moeckel-AP-2009} 
\begin{eqnarray}
	\label{eq:<N>-2}
	N(n,t)=n+(2n+1)\sinh^{2}[r(t)],
\end{eqnarray}
and 
\begin{eqnarray}
	\sigma_{\hat{N}}^{2}(n,t)=\frac{1}{2}(n^{2}+n+1)\sinh^{2}[2r(t)].
\end{eqnarray}
This means that a system can be excited due to the temporal variations in its frequency.
Because of this, the transition probability ${\cal P}(t)_{\mu\to\nu}$ between different states $\mu$ and $\nu$ is given by \cite{Coelho-Entropy-2022}
\begin{eqnarray}
{\cal P}(t)_{\mu\to\nu}=\frac{2^{\mu+\nu}[\min(\mu,\nu)!]^{2}[N(0,t)]^{\frac{|\nu-\mu|}{2}}}{\mu!\nu![N(0,t)+1]^{\frac{1}{2}}}\left|\sum_{q=\frac{|\nu-\mu|}{2}}^{\frac{\mu+\nu}{2}}\frac{\left(\begin{array}{c}
		\frac{\mu+\nu}{2}\\
		q
	\end{array}\right)\left(\begin{array}{c}
		\frac{\mu+\nu+2q-2}{4}\\
		\frac{\mu+\nu}{2}
	\end{array}\right)q!}{\bigl(q-\frac{|\nu-\mu|}{2}\bigr)![N(0,t)+1]^{\frac{q}{2}}}\right|^{2},
\label{eq:Prob-m-n-N}
\end{eqnarray}
for even values of $|\nu-\mu|$ and ${\cal P}(t)_{\mu\to \nu}=0$ for odd values of $|\nu-\mu|$, where $\min(\mu,\nu)$ is the smallest value between $ \mu $ and $ \nu $. 
For a TDHO initially in the fundamental state, the probability of this system to be excited to different energy levels is ${\cal P}_{e}(t)=1-{\cal P}(t)_{0\,\to\,0}$ \cite{Tibaduiza-BJP-2020}.

\section{The model}\label{sec:model}

Let us consider the frequency model defined by
\begin{eqnarray}
	\omega(t)=\begin{cases}
		\omega_{0}, & t<0,\\
		\omega_{\text{int}}(t), & 0\leq t\leq\tau,\\
		\omega_{f}, & t>\tau,
	\end{cases}
	\label{eq:saltos}	
\end{eqnarray}
where $\omega_{0}$ and $\omega_{f}$ are the initial and final frequencies, $ \tau $ is the duration of the time interval separating these frequencies, and $\omega_{\text{int}}(t)$ is a time-dependent intermediary frequency.
Note that $\omega(t)$ does not necessarily have to be continuous at $t=0$ and $t=\tau$, i.e., we can have $\omega_{\text{int}}(t)|_{t=0}\neq\omega_{0}$ and $\omega_{\text{int}}(t)|_{t=\tau}\neq\omega_{f}$.
In this section, we show two approaches for obtaining the parameter $\rho(t)$ associated with model \eqref{eq:saltos}, and discuss the effects caused by frequency changes on the quantities that depend on the squeezing parameter in the interval $t>\tau$.


\subsection{Solution of the parameter $\rho(t)$}\label{sec:sol-rho}

Due to the form of Eq. \eqref{eq:saltos}, the parameter $\rho(t)$ can be written as [hereafter, the same convention of indices and time-intervals is considered for $r(t)$ and all related quantities]: 
\begin{eqnarray}
	\rho(t)=\begin{cases}
		\rho_{0}, & t<0,\\
		\rho_{\text{int}}(t), & 0\leq t\leq\tau,\\
		\rho_{f}(t), & t>\tau,
	\end{cases}
	\label{eq:rho(t)}
\end{eqnarray}
in which $\rho_0$ is given by Eq. \eqref{eq:rho0}, and $\rho_{\text{int}}(t)$ and $\rho_{f}(t)$ are the parameters corresponding to the intermediary and final frequencies, respectively.
A first approach to determine $\rho(t)$ is discussed in Sec. \ref{sec:formal}, and is based on the formal solution of Eq. \eqref{eq:equação de Ermakov-Pinney} \cite{Pinney-PAMS-1950,Carinena-IJGMMP-2009,Lima-JMO-2009,Mancas-AMC-2015}, in which it is assumed that $\omega_\text{int}(t)$ is prescribed.
A second approach, discussed in Sec. \ref{sec:ansatz}, is based on methods to find shortcuts to adiabaticity \cite{Chen-PRL-2010,Beau-Entropy-2020} and consists of defining an ansatz for $\rho(t)$ and,
using Eq. \eqref{eq:equação de Ermakov-Pinney}, determining the correspondent frequency $\omega_\text{int}(t)$. 
The advantage of using one or the other approach will become clear in Sec. \ref{sec:aplicações}. 
Next, we calculate the solutions for $\rho(t)$ using these two approaches. 

\subsubsection{Interval $0\leq t\leq\tau$: formal approach}\label{sec:formal}

For the interval $0\leq t\leq\tau$, the solution based in formal approach for $\rho_{\text{int}}(t)$ is given by \cite{Pinney-PAMS-1950,Carinena-IJGMMP-2009,Lima-JMO-2009,Mancas-AMC-2015}
\begin{eqnarray}	
	\rho_{\text{int}}(t)=\sqrt{A_{\text{int}}u_{\text{int}}(t)^{2}+B_{\text{int}}v_{\text{int}}(t)^{2}+2C_{\text{int}}u_{\text{int}}(t)v_{\text{int}}(t)},
	\label{eq:sol-rho1-formal}
\end{eqnarray}
where $u_\text{int}(t)$ and $v_\text{int}(t)$ are two linearly independent solutions of the classical harmonic oscillator equation:
\begin{eqnarray}
	\label{eq:homogênea}	
	\ddot{z}(t)+\omega_{\text{int}}(t)^{2}z(t)=0,
\end{eqnarray}
and the constants $A_\text{int}$, $B_\text{int}$ and $C_\text{int}$ can be determined from the initial conditions 
\begin{eqnarray}
	\rho_{\text{int}}(t)|_{t=0}=\rho_{0},\;\;\;\dot{\rho}_{\text{int}}(t)|_{t=0}=0,
\end{eqnarray}
and by the relation
\begin{eqnarray}
	A_{\text{int}}B_{\text{int}}-C_{\text{int}}^{2}=\frac{1}{\{m_{0}W[u_{\text{int}}(t),v_{\text{int}}(t)]|_{t=0}\}^{2}},
	\label{eq:A1B1-C1}
\end{eqnarray}
where $W[u_{\text{int}}(t),v_{\text{int}}(t)]=u_{\text{int}}(t)\dot{v}_{\text{int}}(t)-\dot{u}_{\text{int}}(t)v_{\text{int}}(t)$ is the Wronskian determinant of $u_{\text{int}}(t)$ and $v_{\text{int}}(t)$, which results in
\begin{align}
	\label{eq:A1}
	A_{\text{int}}\;=\; & \;\frac{\omega_{0}^{2}v_{\text{int}}(t)^{2}|_{t=0}+\dot{v}_{\text{int}}(t)^{2}|_{t=0}}{m_{0}\omega_{0}\{W[u_{\text{int}}(t),v_{\text{int}}(t)]|_{t=0}\}^{2}},\\
	\label{eq:B1}
	B_{\text{int}}\;=\; & \;\frac{\omega_{0}^{2}u_{\text{int}}(t)^{2}|_{t=0}+\dot{u}_{\text{int}}(t)^{2}|_{t=0}}{m_{0}\omega_{0}\{W[u_{\text{int}}(t),v_{\text{int}}(t)]|_{t=0}\}^{2}},\\
	\label{eq:C1}
	C_{\text{int}}\;=\; & -\frac{\omega_{0}^{2}u_{\text{int}}(t)|_{t=0}v_{\text{int}}(t)|_{t=0}+\dot{u}_{\text{int}}(t)|_{t=0}\dot{v}_{\text{int}}(t)|_{t=0}}{m_{0}\omega_{0}\{W[u_{\text{int}}(t),v_{\text{int}}(t)]|_{t=0}\}^{2}}.
\end{align}
%

\subsubsection{Interval $0\leq t\leq\tau$: ansatz approach}\label{sec:ansatz}

For the interval $0\leq t\leq\tau$, the solution based in ansatz approach for $\rho_{\text{int}}(t)$ is defined by
\begin{eqnarray}	
	\rho_{\text{int}}(t)=\sum_{j=1}^{6}\Gamma_{j}a_{j}(t),
	\label{eq:sol-rho1-ansatz}
\end{eqnarray}
where $a_{j}(t)$ are linearly independent functions, and $\Gamma_{j}$ are constants determined by the following boundary conditions at $t=0$ \cite{Chen-PRL-2010,Beau-Entropy-2020},
\begin{eqnarray}
	\label{eq:cond-t0}
	\rho_{\text{int}}(t)|_{t=0}=\rho_{0},\;\;\;\dot{\rho}_{\text{int}}(t)|_{t=0}=0,\;\;\;\ddot{\rho}_{\text{int}}(t)|_{t=0}=0,
\end{eqnarray}
and, at $t=\tau$, 
\begin{eqnarray}
	\label{eq:cond-tau}
	\rho_{\text{int}}(t)|_{t=\tau}=\delta,\;\;\;\dot{\rho}_{\text{int}}(t)|_{t=\tau}=\epsilon,\;\;\;\ddot{\rho}_{\text{int}}(t)|_{t=\tau}=\gamma.
\end{eqnarray}
Using these conditions, one can find the solutions for $\Gamma_{j}$ in terms of $a_{j}(t)|_{t=0}$ and $a_{j}(t)|_{t=\tau}$, but the expressions are too long to be written here.
After determining $\rho_{\text{int}}(t)$, the strategy is to obtain the frequency $\omega_{\text{int}}(t)$ using the Ermakov-Pinney equation \eqref{eq:equação de Ermakov-Pinney}:
\begin{eqnarray}
	\omega_{\text{int}}(t)=\sqrt{\frac{1}{m_{0}^{2}\rho_{\text{int}}(t)^{4}}-\frac{\ddot{\rho}_{\text{int}}(t)}{\rho_{\text{int}}(t)}}.
	\label{eq:w-in}
\end{eqnarray}
As will be seen later, the presence of the term $\epsilon$ in Eq. \eqref{eq:cond-tau} is essential for the generation of squeezing after $t>\tau$ (unlike what happens in the context of shortcuts to adiabaticity, where $\epsilon=0$, which naturally leads to a null final squeezing).  
Thus, the techniques usually used in the context of shortcuts to adiabaticity are extended to set up the parameter $\rho(t)$ [solution of the Ermakov-Pinney equation \eqref{eq:equação de Ermakov-Pinney}] and, consequently, the intermediate frequency $\omega_{\text{int}}(t)$, to produce protocols with non-zero final squeezing. 
This allows us to control, for different intermediate frequencies, the value that $r_{f}$ can take on.
In summary, the squeezing equivalence found in protocols for shortcuts to adiabaticity
(leading to $r_{f}=0$) is a particular case of the squeezing equivalence found in more general protocols leading to $r_{f}>0$.
The term $\gamma$, on the other hand, is associated with continuous or discontinuous frequency transitions at $t=\tau$ [if $\gamma=0$, we have $\omega_{\text{int}}(t)|_{t=\tau}=\omega_{f}$, otherwise, $\omega_{\text{int}}(t)|_{t=\tau}\neq\omega_{f}$].
%


\subsubsection{Interval $t>\tau$}\label{Sec:sol-t>tau}

For the interval $t>\tau$, the general solution for $\rho_f(t)$ has the form
\begin{eqnarray}	
	\rho_{f}(t)=\sqrt{A_{f}u_{f}(t)^{2}+B_{f}v_{f}(t)^{2}+2C_{f}u_{f}(t)v_{f}(t)},
	\label{eq:sol-rho2-genérico}
\end{eqnarray}
where $u_{f}(t)=\sin(\omega_{f}t)$ and $v_{f}(t)=\cos(\omega_{f}t)$.
By means of the continuity conditions 
\begin{eqnarray}
	\rho_{\text{int}}(t)|_{t=\tau}=\rho_{f}(t)|_{t=\tau},\;\;\;\dot{\rho}_{\text{int}}(t)|_{t=\tau}=\dot{\rho}_{f}(t)|_{t=\tau},
\end{eqnarray}
and the relation 
\begin{eqnarray}
	A_{f}B_{f}-C_{f}^{2}=\frac{1}{(m_{0}\omega_{f})^{2}},
	\label{eq:A2B2-C2}
\end{eqnarray}
one can express the coefficients $A_f$, $B_f$ and $C_f$ in terms of the parameter $\rho_{\text{int}}(t)$ and its time derivative $\dot{\rho}_{\text{int}}(t)$, both calculated at $t=\tau$ [see Eq. \eqref{eq:cond-tau}]:
\begin{align}
	\label{eq:A2-genérico}
	A_{f}\;= & \,\;\biggl(\frac{\epsilon^{2}}{\omega_{f}^{2}}+\frac{1}{m_{0}^{2}\omega_{f}^{2}\delta^{2}}-\delta^{2}\biggr)\cos^{2}(\omega_{f}\tau)+\delta^{2}+\frac{\epsilon\delta\sin(2\omega_{f}\tau)}{\omega_{f}},\\
	\label{eq:B2-genérico}
	B_{f}\;= & \;\,\biggl(\frac{\epsilon^{2}}{\omega_{f}^{2}}+\frac{1}{m_{0}^{2}\omega_{f}^{2}\delta^{2}}-\delta^{2}\biggr)\sin^{2}(\omega_{f}\tau)+\delta^{2}-\frac{\epsilon\delta\sin(2\omega_{f}\tau)}{\omega_{f}},\\
	C_{f}\;= & \;\,\frac{[m_{0}^{2}(\omega_{f}^{2}\delta^{4}-\epsilon^{2}\delta^{2})-1]\sin(2\omega_{f}\tau)}{2m_{0}^{2}\omega_{f}^{2}\delta^{2}}+\frac{\epsilon\delta\cos(2\omega_{f}\tau)}{\omega_{f}}.
	\label{eq:C2-genérico}
\end{align}
Effectively, obtaining $\rho_{f}(t)$ comes down to determining the constants $\delta$ and $\epsilon$.

\subsection{Squeezing parameter and related quantities for $t>\tau$}\label{sec:relat-quant}

It is interesting to analyze the consequences generated by the intermediary frequency $\omega_{\text{int}}(t)$ on the dynamics of the system for $t>\tau$.
From Eqs. \eqref{eq:r}, \eqref{eq:lambda(t)} and \eqref{eq:sol-rho2-genérico}-\eqref{eq:C2-genérico}, we have that 
\begin{eqnarray}
	r_{f}=\cosh^{-1}[(\lambda_{f})^{\frac{1}{2}}],
	\label{eq:rf-w}
\end{eqnarray}
where
\begin{eqnarray}
	\lambda_{f}=\frac{m_{0}^{2}\epsilon^{2}+\delta^{-2}+m_{0}^{2}\omega_{f}^{2}\delta^{2}+2m_{0}\omega_{f}}{4m_{0}\omega_{f}}.
	\label{eq:lambda}
\end{eqnarray}
Note that the squeezing parameter is time-independent for $t>\tau$, because it only depends on $\delta$ and $\epsilon$ (ignoring the terms $m_0$ and $\omega_f$, which we assume are prescribed).
In other words, it is enough to know the values of $\rho_{\text{int}}(t)$ and $\dot{\rho}_{\text{int}}(t)$ in $t=\tau$ to determine the dynamics of the system after the instant $\tau$.
Therefore, all quantities that depend only on $r_{f}$ will be time-independent for $t>\tau$, where the frequency of the TDHO is $\omega_f$.
Besides this, the squeezing phase $\phi_{f}(t)$ is given by
\begin{eqnarray}
\phi_{f}(t)=\cos^{-1}\biggl\{\frac{\sech(r_{f})}{4m_{0}\omega_{f}\sinh(r_{f})}\bigl\{2m_{0}^{2}\omega_{f}\epsilon\delta\sin[2\omega_{f}(t-\tau)]+[m_{0}^{2}(\omega_{f}^{2}\delta^{2}\;-\epsilon^{2})-\delta^{-2}]\cos[2\omega_{f}(t-\tau)]\bigr\}\biggr\}.
\label{eq:phif-w}
\end{eqnarray}
As can be seen, in contrast with $r_{f}$, $\phi_{f}(t)$ is time-dependent. 
Consequently, $\sigma_{\hat{x}}^{2}(n,t)$ and $\sigma_{\hat{p}}^{2}(n,t)$ will also depend explicitly on time for $t>\tau$, because these quantities depend on $\phi_{f}(t)$.

Furthermore, we obtain the following expression for the mean energy $E_{f}(n)$:  
\begin{eqnarray}
	\label{eq:E2-f(t)}
	E_{f}(n)=\cosh(2r_{f})\overline{E}_{f}(n),
\end{eqnarray}
where $\overline{E}_{f}(n)=(n+1/2)\hbar\omega_{f}$.
Notably, the mean energy of the system for $t>\tau$ is time-independent, as expected.  
However, it can be seen that $E_{f}(n) \geq \overline{E}_{f}(n)$, even when the initial and final frequencies are the same ($\omega_{f}=\omega_{0} $) \cite{Coelho-Entropy-2022}. 
The argument is similar for $\sigma_{\hat{H}}^{2}(n,t)$ in the interval $t>\tau$, where $\omega(t)=\omega_{f}$ and $r(t)=r_{f}$ [see Eq. \eqref{eq:var-E}].
Using Eqs. \eqref{eq:<N>-2}, \eqref{eq:rf-w}, \eqref{eq:lambda} and \eqref{eq:E2-f(t)}, one can determine the mean number of excitations $N_{f}(n)$ as
\begin{eqnarray}
	\label{eq:E_2-f(t)-N_2-f(t)}	
	N_{f}(n)=\frac{E_{f}(n)-\overline{E}_{f}(0)}{2\overline{E}_{f}(0)}.
\end{eqnarray}
We highlight that there is excitation even for $ n=0 $ since $E_{f}(0)>\overline{E}_{f}(0)$, which means that, under modulations in its frequency, a TDHO initially in its ground state can become excited.
Given this, the transition probability between different states, after the instant $\tau$, is obtained by substituting Eq. \eqref{eq:E_2-f(t)-N_2-f(t)} (with $n=0$) into Eq. \eqref{eq:Prob-m-n-N}.
So, even when the frequency returns to its initial value ($\omega_{f}=\omega_{0} $) after an instant $ \tau $, one can find a non-zero $ \mu\to\nu $ transition probability. This is a direct consequence of Eqs. \eqref{eq:E2-f(t)} and \eqref{eq:E_2-f(t)-N_2-f(t)}.
The Eqs. \eqref{eq:rf-w}-\eqref{eq:E_2-f(t)-N_2-f(t)} generalize, for a generic intermediary frequency $\omega_{\text{int}}(t)$ and a final frequency $\omega_{f}$, the results found in Ref. \cite{Coelho-Entropy-2022} for the particular case where $\omega_{\text{int}}(t)=\omega_{1}$ and $\omega_{f}=\omega_{0}$.

\section{General condition for squeezing equivalence of quantum harmonic oscillators under different frequency modulations}\label{sec:equivalence}

In this section, we discuss how to configure a set of TDHOs, subject to different intermediary frequency modulations (discontinuous or continuous), so that they have the same squeezing parameter $r_f$. 
We also discuss the physical implications of this equivalence.

Let us consider another TDHO with frequency $\omega^{\prime}(t)$, defined by
\begin{eqnarray}
	\omega^{\prime}(t)=\begin{cases}
		\omega_{0}, & t<0,\\
		\omega_{\text{int}}^{\prime}(t), & 0\leq t\leq\tau,\\
		\omega_{f}^{\prime}, & t>\tau.
	\end{cases}
	\label{eq:saltos-2}
\end{eqnarray}
The associated parameter $\rho^{\prime}(t)$ is given by
\begin{eqnarray}
	\rho^{\prime}(t)=\begin{cases}
		\rho_{0}, & t<0,\\
		\rho_{\text{int}}^{\prime}(t), & 0\leq t\leq\tau,\\
		\rho_{f}^{\prime}(t), & t>\tau.
	\end{cases}
\end{eqnarray}
The solutions for $\rho_{\text{int}}^{\prime}(t)$ and $\rho_{f}^{\prime}(t)$ are obtained following the same steps mentioned in Sec. \ref{sec:sol-rho}.
We represent these solutions by inserting a prime superscript in all objects with index “int” or $f$ in Eqs. \eqref{eq:sol-rho1-formal} to \eqref{eq:C2-genérico}.
Thus, following this convention, the squeezing parameter $r_{f}^{\prime}$ can be written by substituting $\omega_{f}\to\omega_{f}^{\prime}$, $\delta\to\delta^{\prime}$ and $\epsilon\to\epsilon^{\prime}$ into Eqs. \eqref{eq:rf-w} and \eqref{eq:lambda}, leading to
\begin{eqnarray}
	r_{f}^{\prime}=\cosh^{-1}[(\lambda_{f}^{\prime})^{\frac{1}{2}}],
	\label{eq:rf'-w'}
\end{eqnarray}
in which
\begin{eqnarray}
	\lambda^{\prime}_{f}=\frac{m_{0}^{2}\epsilon^{\prime2}+\delta^{\prime-2}+m_{0}^{2}\omega_{f}^{\prime2}\delta^{\prime2}+2m_{0}\omega_{f}^{\prime}}{4m_{0}\omega_{f}^{\prime}}.
	\label{eq:lambda'}
\end{eqnarray}
Given that, in general, $\omega_{f}\neq \omega_{f}^{\prime}$, $\delta\neq\delta^{\prime}$ and $\epsilon\neq\epsilon^{\prime}$, then $r_{f}\neq r_{f}^{\prime}$.
On the other hand, even when $\omega_{\text{int}}(t)\neq\omega_{\text{int}}^{\prime}(t)$, it is possible to determine a set of conditions which, if met, result in $r_{f}=r_{f}^{\prime}$.
Comparing Eqs. \eqref{eq:rf-w}, \eqref{eq:lambda}, \eqref{eq:rf'-w'} and \eqref{eq:lambda'}, it can be seen that the squeezing equivalence is achieved when
\begin{eqnarray}
	\omega_{f}=\omega_{f}^{\prime},\;\;\;\delta=\delta^{\prime},\;\;\;\epsilon=\epsilon^{\prime}.
	\label{eq:cond-equi}
\end{eqnarray}
Thus, Eq. \eqref{eq:cond-equi} shows how to configure two TDHOs, with different intermediary frequency modulations, so that they result in a same value for the squeezing parameter or, in other words, in a squeezing equivalence. 

As can be seen, the squeezing equivalence depends on the boundary conditions that $\rho(t)$, $\rho^{\prime}(t)$, $\dot{\rho}(t)$ and $\dot{\rho}^{\prime}(t)$ satisfy at $t=\tau$.
From Eqs. \eqref{eq:Psi}-\eqref{eq:phi_{n}}, one can see that these parameters determine
the behavior of the wave functions $\Psi(x,t)$ and $\Psi^{\prime}(x,t)$
\cite{Huang-Chaos-2020,Anderson-JO-1994}, so that the squeezing equivalence is a consequence 
from that $\rho_{f}(t)=\rho_{f}^{\prime}(t)$ and $\dot{\rho}_{f}(t)=\dot{\rho}_{f}^{\prime}(t)$ (see Sec. \ref{Sec:sol-t>tau}).
It is worth noting that $\Psi(x,t)$ and $\Psi^{\prime}(x,t)$ differ by a phase factor
[$\alpha_{n}(t)\neq\alpha_{n}^{\prime}(t)$] in the interval $t>\tau$ [see Eq. \eqref{eq:alpha_{n}}].
However, these phase factors do not contribute to the physical quantities investigated here.
This, in turn, means that all physical quantities investigated here, depending on the squeezing parameter $r_f$ (see Sec. \ref{sec:relat-quant}) and, consequently, on the parameters $\rho(t)$ and $\dot{\rho}(t)$, will be the same for $t>\tau$.

In summary, two TDHOs with different intermediary frequencies [$\omega_{\text{int}}(t)\neq\omega_{\text{int}}^{\prime}(t)$]
can have be configured [Eq. \eqref{eq:cond-equi}] in such a way that
they exhibit the same final squeezing parameter ($r_{f}=r_{f}^{\prime}\geq 0$).
This is defined here a squeezing equivalence between these
two TDHOs.
We highlight that the criteria of squeezing equivalence proposed here
implies in the invariance of physical properties of the TDHOs,
in such a way that all physical quantities depending on the squeezing parameter will be the same for $t>\tau$ (as, for instance, quantum fluctuations of the
position and momentum operators, quantum fluctuations of the energy, mean number of excitations and its fluctuations, and the transition probabilities between different states).
As mentioned before, a particular case of this squeezing equivalence 
($r_{f}=r_{f}^{\prime}=0$) is found, for example, in Refs. \cite{Janszky-PRA-1992,Chen-PRL-2010},
whereas here we are discussing a more general case ($r_{f}=r_{f}^{\prime}\geq0$).

The extension of this squeezing equivalence to two different TDHOs with $k$ intermediary frequency modulations is natural. 
To do this, simply consider that the time-dependent intermediary frequency $\omega_{\text{int}}(t)$ in Eq. \eqref{eq:saltos} is given by 
\begin{eqnarray}
	\label{eq:w-int-N}	
	\omega_{\text{int}}(t)=\begin{cases}
		\omega_{1}(t), & 0\leq t<t_{1},\\
		\omega_{2}(t), & t_{1}\leq t<t_{2},\\
		\vdots & \vdots\\
		\omega_{k}(t), & t_{k-1}\leq t\leq\tau,
	\end{cases}
\end{eqnarray}
where $k\geq1$. 
For $t<0$ and $t>\tau$, the angular frequency remains $\omega_{0}$ and $\omega_{f}$, respectively.
The parameter $\rho_\text{int}(t)$ associated with Eq. \eqref{eq:w-int-N} is written as 
\begin{eqnarray}
	\label{eq:rho-int-N}	
	\rho_{\text{int}}(t)=\begin{cases}
		\rho_{1}(t), & 0\leq t<t_{1},\\
		\rho_{2}(t), & t_{1}\leq t<t_{2},\\
		\vdots & \vdots\\
		\rho_{k}(t), & t_{k-1}\leq t\leq\tau,
	\end{cases}
\end{eqnarray}
and is constructed in the same way as already discussed.
Thus, for squeezing equivalence to occur, defining $\delta_{k}=\rho_{k}(t)|_{t=\tau}$ and $\epsilon_{k}=\dot{\rho}_{k}(t)|_{t=\tau}$, it is sufficient that 
\begin{eqnarray}
	\label{eq:equi-N}	
	\omega_{f}=\omega_{f}^{\prime},\;\;\;\delta_{k}=\delta_{k}^{\prime},\;\;\;\epsilon_{k}=\epsilon_{k}^{\prime}.
\end{eqnarray}
Moreover, the generalization of the squeezing equivalence to several different TDHOs with $k$ intermediary frequency modulations is also direct, it is enough that  
\begin{eqnarray}
	\label{eq:wf=wf'=...wf^k}	
	\omega_{f}=\omega_{f}^{\prime}=\omega_{f}^{\prime\prime}=\omega_{f}^{\prime\prime\prime}=\ldots,
\end{eqnarray}
as well as
\begin{eqnarray}
	\label{eq:deltaf=deltaf'=...deltaf^k}	
	\delta_{k}=\delta_{k}^{\prime}=\delta_{k}^{\prime\prime}=\delta_{k}^{\prime\prime\prime}=\ldots,
\end{eqnarray}
and
\begin{eqnarray}
	\label{eq:epsilonf=epsilonf'=...epsilonf^k}	
	\epsilon_{k}=\epsilon_{k}^{\prime}=\epsilon_{k}^{\prime\prime}=\epsilon_{k}^{\prime\prime\prime}=\ldots,
\end{eqnarray}
in which “$\ldots$” indicates the several different TDHOs.

\section{Applications}\label{sec:aplicações}

Now, we will find solutions for Eq. \eqref{eq:cond-equi} [or its variants \eqref{eq:equi-N}-\eqref{eq:epsilonf=epsilonf'=...epsilonf^k}], for specific models of
TDHOs, via the approaches discussed in Secs. \ref{sec:formal} and \ref{sec:ansatz}. 
The approach in Sec. \ref{sec:formal} is useful when $r_{f}$ is not prescribed and the interest is in controlling the form of intermediary frequency modulations of different TDHOs, so that they are previously prescribed.  
In Table \ref{tabela}, we indicate such prescription as a ``semi-prescription'', in the sense that these frequencies in fact also depend on $\tau$, which is not prescribed and must be determined later.
In this way, once these frequencies are described, Eq. \eqref{eq:cond-equi} [or its variants \eqref{eq:equi-N}-\eqref{eq:epsilonf=epsilonf'=...epsilonf^k}] leads to numerical calculations (in general) to obtain $r_{f}$ and $\tau$.
In summary, we depart from that the general forms of the intermediary frequencies are prescribed, and impose that the conditions \eqref{eq:cond-equi} [or its variants \eqref{eq:equi-N}-\eqref{eq:epsilonf=epsilonf'=...epsilonf^k}] are satisfied, which, in turn, enable us to find numerically the exact behaviors of the frequencies of the TDHOs, and also the values of $r_{f}$ and $\tau$.
In contrast, when the interest is not in previously prescribing the form of the intermediary frequencies, but rather $r_f$ and $\tau$ (and also some constraints, such as frequency limitations and minimization of energy), the approach to be used is the one discussed in Sec. \ref{sec:ansatz}. 
In this way, we start imposing that the conditions \eqref{eq:cond-equi} [or its variants \eqref{eq:equi-N}-\eqref{eq:epsilonf=epsilonf'=...epsilonf^k}] are satisfied, and this, in turn, enable us to find analytically the intermediary frequencies.
The outline of a comparison between prescribed and non-prescribed parameters in these two approaches are illustrated in Table \ref{tabela}.	
\begin{table}[h!]
	\centering	
	\begin{tabular}{c|c|c|c|c|c|c|}
		\cline{2-7}
		& $\omega_0$ & $r_0$ & $\omega_f$ & $r_f$ & $\omega_{\text{int}}(t)$ & $\tau$ \\ \hline
		\multicolumn{1}{|c|}{Formal approach} & $P$        & $P$   & $P$        & $N$   & $S$                      & $N$    \\ \hline
		\multicolumn{1}{|c|}{Ansatz approach} & $P$        & $P$   & $P$        & $P$   & $N$                      & $P$    \\ \hline
	\end{tabular}
	\caption{Table summarizing prescribed ($P$), non-prescribed ($N$), and semi-prescribed ($S$) parameters in the formal approach (Sec. \ref{sec:formal}), and ansatz approach (Sec. \ref{sec:ansatz}).}
	\label{tabela}	
\end{table}

In the following, using the approach of Sec. \ref{sec:formal}, we discuss how the equivalence found by Janszky and Adam \cite{Janszky-PRA-1992} is a particular case of the squeezing equivalence found by us (Sec. \ref{Sec:Janszky-Adam}). 
Moreover, through the approach used in Sec. \ref{sec:ansatz}, we show the relationship between this squeezing equivalence and the protocols that seek shortcuts to adiabaticity (Sec. \ref{sec:SE and SA}).
Finally, we also analyze the squeezing equivalence between generic frequency modulations and the frequency that minimizes the intermediary mean energy (Sec. \ref{sec:se-energy}).	

\subsection{Recovering the squeezing equivalence found by Janszky and Adam}\label{Sec:Janszky-Adam}

The squeezing equivalence found by Janszky and Adam in Ref. \cite{Janszky-PRA-1992} can be obtained as follows.
Firstly, considering $\omega_{\text{int}}(t)=\omega_{1}$ and $\omega_{f}=\omega_{0}$ in Eq. \eqref{eq:saltos}, as well as $\omega_{\text{int}}^{\prime}(t)=\omega_{0}$ and $\omega_{f}^{\prime}=\omega_{0}$ in Eq. \eqref{eq:saltos-2}, we have
\begin{eqnarray}
\omega(t)=\begin{cases}
		\omega_{0}, & t<0,\\
		\omega_{1}, & 0\leq t\leq\tau,\\
		\omega_{0}, & t>\tau,
\end{cases}
\label{eq:Janszky-1}
\end{eqnarray} 
and
\begin{eqnarray}
\omega^{\prime}(t)=\begin{cases}
	\omega_{0}, & t<0,\\
	\omega_{0}, & 0\leq t\leq\tau,\\
	\omega_{0}, & t>\tau.
\end{cases}
\label{eq:Janszky-2}	
\end{eqnarray}
Note that in Ref. \cite{Janszky-PRA-1992} the intermediary frequency is prescribed, in such way it corresponds to sudden frequency jumps, intending to generate squeezed states. In this way, the intermediary frequency for $\omega(t)$ in Eq. \eqref{eq:Janszky-1} is defined unless $\tau$, so that it is semi-prescribed. Moreover, note that $r_{f}$ and $\tau$ are not prescribed, but to be determined later.

From Eq. \eqref{eq:cond-equi}, one can see that Eqs. \eqref{eq:Janszky-1} and \eqref{eq:Janszky-2} satisfy the condition $\omega_{f}=\omega_{f}^{\prime}$.
To determine the conditions that satisfy $\delta=\delta^{\prime}$ and $\epsilon=\epsilon^{\prime}$, we need to know $\rho_{\text{int}}(t)$ and $\rho_{\text{int}}^{\prime}(t)$, which for this case are given respectively by \cite{Coelho-Entropy-2022}
\begin{eqnarray}
	\rho_{\text{int}}(t)=\sqrt{\frac{\omega_{0}\sin^{2}(\omega_{1}t)}{m_{0}\omega_{1}^{2}}+\frac{\cos^{2}(\omega_{1}t)}{m_{0}\omega_{0}}},\;\;\;\rho_{\text{int}}^{\prime}(t)=\rho_{0}.
\end{eqnarray}
In this case, the only way these conditions can be satisfied (beyond the trivial case $\omega_{1}=\omega_{0}$) is if $\tau=s\pi/\omega_1$ ($s\in\mathbb{N}$).
Consequently, for $\tau=s\pi/\omega_1$, we find $\delta=\delta^{\prime}=\rho_{0}$ and $\epsilon=\epsilon^{\prime}=0$, which implies $r_{f}=r^{\prime}_{f}=0$. 
This is the squeezing equivalence found by Janszky and Adam in Ref. \cite{Janszky-PRA-1992}. 
For more details on this model (apart from the squeezing equivalence) see, for instance, Refs. \cite{Tibaduiza-BJP-2020,Coelho-Entropy-2022}.

\begin{figure}[h!]
	\centering
	\subfigure[\label{fig:rho-J-M}]{\epsfig{file=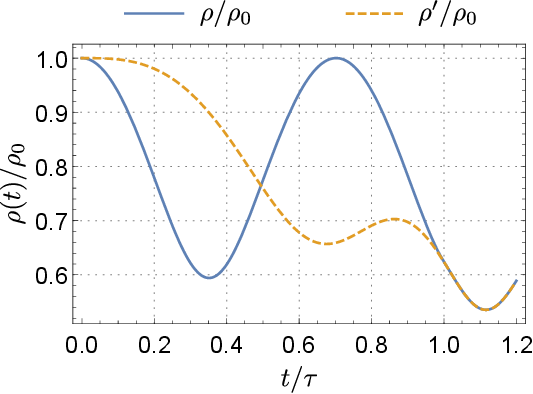,  width=0.35 \linewidth}}
	\hspace{1.0mm}
	\centering
	\subfigure[\label{fig:omega-J-M}]{\epsfig{file=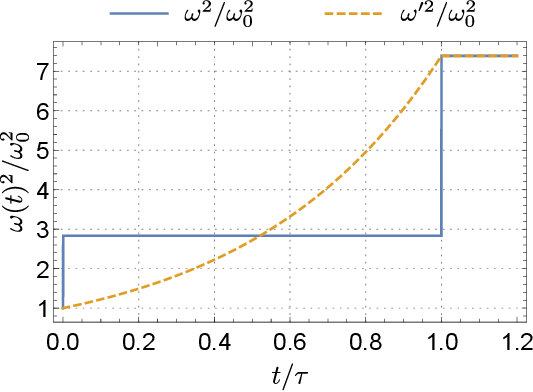, width=0.35 \linewidth}}
	\hspace{1.0mm}
	\centering
	\subfigure[\label{fig:r-J-M}]{\epsfig{file=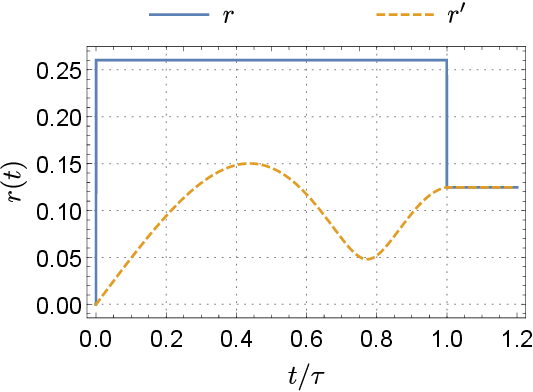, width=0.38 \linewidth}}
	\caption{Comparison between (a) $\rho(t)/\rho_{0}$ and $\rho^{\prime}(t)/\rho_{0}$, (b) $\omega(t)^{2}/\omega_{0}^{2}$ and $\omega^{\prime}(t)^{2}/\omega_{0}^{2}$, and (c) $r(t)$ and $r^{\prime}(t)$, as a function of $t/\tau$, for the case where $\omega(t)$ and $\omega^\prime(t)$ are given by Eqs. \eqref{eq:w-num-1} and \eqref{eq:w-num-2}. 
		The unprimed quantities are represented by the solid lines, while the primed ones by the dashed lines.
		In these figures we consider $\omega_{0}\tau=2.657887$ and $\omega_{1}\tau=4.473165$.}
	\label{fig:rho-omega-r-J-M}
\end{figure}

As can be seen, obtaining the conditions that lead to squeezing equivalence for the case studied above is relatively simple. 
However, for more complicated models, it is not trivial to find these conditions. To exemplify this, let us consider
\begin{eqnarray}
	\label{eq:w-num-1}	
	\omega(t)=\begin{cases}
		\omega_{0}, & t<0,\\
		\omega_{1}, & 0\leq t\leq\tau,\\
		\omega_{0}e, & t>\tau,
	\end{cases}
\end{eqnarray}
and
\begin{eqnarray}
	\label{eq:w-num-2}
	\omega^{\prime}(t)=\begin{cases}
		\omega_{0}, & t<0,\\
		\omega_{0}e^{t/\tau}, & 0\leq t\leq\tau,\\
		\omega_{0}e, & t>\tau.
	\end{cases}
\end{eqnarray}
In Refs. \cite{Alsing-PRL-2005,Menicucci-PRA-2007}, a similar exponential frequency is used to show how an ion trap can be used to simulate quantum fields in an expanding universe.
In this case, the squeezing equivalence conditions can only be obtained numerically, because Eq. \eqref{eq:cond-equi} forms a set of transcendental equations [for this example it can be shown, from the approach developed in Sec. \ref{sec:formal}, that $\rho_{\text{int}}^{\prime}(t)$ will depend on Bessel functions].
One possible solution is given by $\omega_{0}\tau=2.657887$ and $\omega_{1}\tau=4.473165$, and, in Fig. \ref{fig:rho-omega-r-J-M}, we show for this solution the time evolution of $\rho(t)/\rho_{0}$ and $\rho^{\prime}(t)/\rho_{0}$ [Fig. \ref{fig:rho-J-M}], $\omega(t)^{2}/\omega_{0}^{2}$ and $\omega^{\prime}(t)^{2}/\omega_{0}^{2}$ [Fig. \ref{fig:omega-J-M}], as well as $r(t)$ and $r^{\prime}(t)$ [Fig. \ref{fig:r-J-M}].
From Figs. \ref{fig:rho-J-M} and \ref{fig:omega-J-M}, one notes that although the parameters $\rho(t)$ and $\rho^{\prime}(t)$, as well as $\omega(t)$ and $\omega^{\prime}(t)$, differ in the interval $0\leq t\leq\tau$, after the instant $\tau$, they become equal.
This leads to $r(t)=r^{\prime}(t)$ for $t>\tau$, as shown in Fig. \ref{fig:r-J-M}, indicating the squeezing equivalence between the TDHOs.

\subsection{Relationship between squeezing equivalence and shortcuts to adiabaticity} \label{sec:SE and SA}

In the context of shortcuts to adiabaticity, TDHOs are utilized to describe the fast frictionless cooling of atoms and quantum gases in harmonic traps \cite{Salamon-PCCP-2009, Chen-PRL-2010,Chen-PRA-2010,Stefanatos-PRA-2010, Choi-PRA-2012,Choi-PRA-2013,Odelin-RMP-2019,Beau-Entropy-2020, Huang-Chaos-2020,Dupays-PRR-2021}. 
The measure of the degree of nonadiabaticity, given by $Q^{*}(t)$, can be written in terms of the squeezing parameter $r(t)$ by $Q^{*}(t)=\cosh[2r(t)]$ \cite{Galve-PRA-2009}.
In our case, we can define the initial and final values of $Q^{*}(t)$ as given by $Q_{0}^{*}=Q^{*}(t<0)$ and $Q_{f}^{*}=Q^{*}(t>\tau)$, respectively. Thus, for nonadiabatic transformations one has $Q_{0}^{*}=1$ and $Q_{f}^{*}>1$, whereas for adiabatic transformations, $Q_{0}^{*}=Q_{f}^{*}=1$ \cite{Galve-PRA-2009}.
This latter condition implies that $r_0=r_f=0$, which leads to null fluctuations in the energy [Eq. \eqref{eq:var-E}], as well as a mean number of excitations that does not change [Eq. \eqref{eq:<N>-2}] [consequently, $	\sigma_{\hat{N}}^{2}(n,t)=0$)], so that the initial and final populations will be the same, as required by protocols that seek shortcuts to adiabaticity \cite{Chen-PRL-2010,Beau-Entropy-2020}.
In this way, a null squeezing parameter in the intervals $t<0$ and $t>\tau$ are the key step to reach protocols that lead to shortcuts to adiabaticity \cite{Choi-PRA-2012,Choi-PRA-2013}.

In this scenario, considering, for example, two different TDHOs, one can see that by making $\delta=\delta^{\prime}=1/(m_{0}\omega_{f})^{\frac{1}{2}}$ and $\epsilon=\epsilon^{\prime}=0$ in Eq. \eqref{eq:cond-equi}, in addition to obtaining the same squeezing parameter $r_{f}=r_{f}^{\prime}=0$ [see Eqs. \eqref{eq:rf-w}, \eqref{eq:lambda}, \eqref{eq:rf'-w'} and \eqref{eq:lambda'}], we also obtain $Q_{f}^{*}=Q_{f}^{*\prime}=1$, i.e., two adiabatic protocols.
Thus, this means that different protocols that seek shortcuts to adiabaticity share the same final squeezing parameters, or in other words, there is an squeezing equivalence between two different adiabatic protocols. 
These ideas can also be extended to $k$ different intermediate frequency modulations and several different TDHOs by using Eqs. \eqref{eq:equi-N}-\eqref{eq:epsilonf=epsilonf'=...epsilonf^k}.
Consequently, the squeezing equivalence (in this case, a null equivalence) can also be used as a way of obtaining different trajectories that result in adiabatic protocols.

To illustrate the ideas mentioned above, let us start assuming that  $\omega(t)$ and $\omega^{\prime}(t)$ are given by Eqs. \eqref{eq:saltos} and \eqref{eq:saltos-2}.
Note that the intermediary frequencies are not prescribed in this case.
In this way, let us consider for $\rho_{\text{int}}(t)$ and $\rho_{\text{int}}^{\prime}(t)$ the following ansatz:
\begin{eqnarray}
	\rho_{\text{int}}(t)=\sum_{j=1}^{6}\Gamma_{j}a_{j}(t),\;\;\;a_{j}(t)=(1+\beta t)^{j-1},
	\label{eq:ansatz polinomial}
\end{eqnarray}
and 
\begin{eqnarray}
	\rho_{\text{int}}^{\prime}(t)=\sum_{j=1}^{6}\Gamma_{j}^{\prime}a_{j}^{\prime}(t),\;\;\;a_{j}^{\prime}(t)=\exp(j\kappa t),
	\label{eq:ansatz exponencial}
\end{eqnarray}
in which $\beta$ and $\kappa$ are parameters with frequency dimension, and $\Gamma_{j}$ and $\Gamma_{j}^{\prime}$ are written in terms of $a_{j}(t)$ and $a_{j}^{\prime}(t)$, respectively, calculated at $t=0$ and $t=\tau$ (see Sec. \ref{sec:ansatz}).
For simplicity, let us also consider a continuous transition in the frequencies at $t=\tau$, i.e., $\omega_{\text{int}}(t)|_{t=\tau}=\omega_{f}$, as well as $\omega_{\text{int}}^{\prime}(t)|_{t=\tau}=\omega_{f}^{\prime}$, which implies
\begin{eqnarray}
	\delta=\frac{1}{\sqrt{m_{0}\omega_{f}}},\;\;\;\delta^{\prime}=\frac{1}{\sqrt{m_{0}\omega_{f}^{\prime}}},\;\;\;\gamma=\gamma^{\prime}=0.
	\label{eq:w-in=wf}
\end{eqnarray}
We remark that the ansatz given in Eq. \eqref{eq:ansatz polinomial} reproduces, for $\beta=1$ and $\epsilon=0$, the same scaling factor $b_{\text{int}}(t)=\rho_{\text{int}}(t)/\rho_{0}$ found in Ref. \cite{Chen-PRL-2010} for a polynomial ansatz.
Moreover, although the intermediary frequencies are not prescribed, in the ansatz approach we can control the values of $r_{f}$ and $\tau$.
Thus, to obtain the squeezing equivalence, one imposes $r_{f}=r_{f}^{\prime}$, which naturally imposes Eq. \eqref{eq:cond-equi} to be satisfied.

\begin{figure}[h!]
	\centering
	\subfigure[\label{fig:rho-P-E}]{\epsfig{file=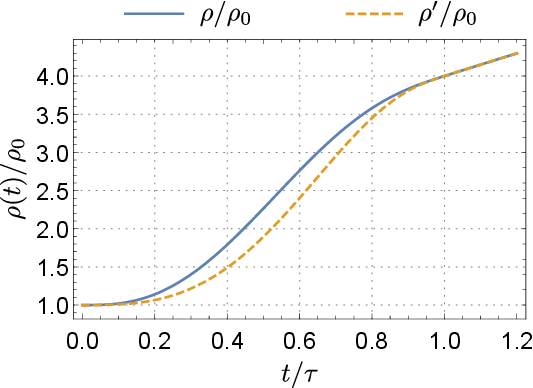,  width=0.35 \linewidth}}
	\hspace{1.0mm}
	\centering
	\subfigure[\label{fig:omega-P-E}]{\epsfig{file=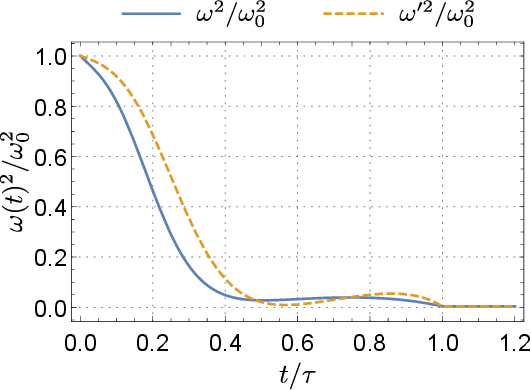, width=0.35 \linewidth}}
	\hspace{1.0mm}
	\centering
	\subfigure[\label{fig:r-P-E}]{\epsfig{file=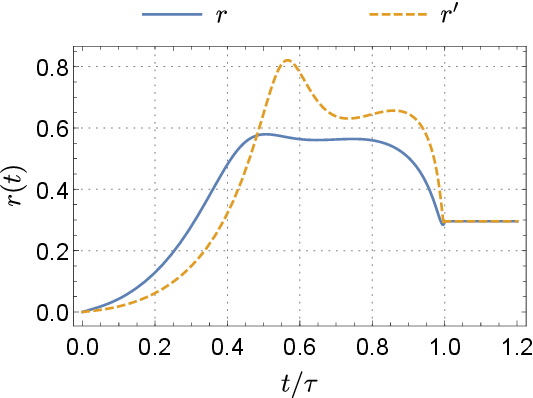, width=0.38 \linewidth}}
	\caption{Comparison between (a) $\rho(t)/\rho_{0}$ and $\rho^{\prime}(t)/\rho_{0}$, (b) $\omega(t)^{2}/\omega_{0}^{2}$ and $\omega^{\prime}(t)^{2}/\omega_{0}^{2}$, and (c) $r(t)$ and $r^{\prime}(t)$, as a function of $t/\tau$, for the case where $\rho_{\text{int}}(t)$ and $\rho_{\text{int}}^\prime(t)$ are given by Eqs. \eqref{eq:ansatz polinomial} and \eqref{eq:ansatz exponencial}, respectively. 
		The unprimed quantities are represented by the solid lines, while the primed ones by the dashed lines.
		In these figures we consider, in arbitrary units, $\omega_{0}=20$, $\omega_{f}=\omega_{f}^{\prime}=\omega_{0}/16$, $\tau=10/\omega_{0}$, $\beta=\kappa=1$, $\delta=\delta^{\prime}=1/(m_{0}\omega_{f})^{\frac{1}{2}}$, $\epsilon=\epsilon^{\prime}=3$ and $\gamma=\gamma^{\prime}=0$.}	
	\label{fig:rho-omega-r-P-E}
\end{figure}

Considering an example where $\epsilon=\epsilon^{\prime}\neq0$
(a nonadiabatic protocol), 
we show in Fig. \ref{fig:rho-omega-r-P-E} the time evolution of $\rho(t)/\rho_{0}$ and $\rho^{\prime}(t)/\rho_{0}$ [Fig. \ref{fig:rho-P-E}], $\omega(t)^{2}/\omega_{0}^{2}$ and $\omega^{\prime}(t)^{2}/\omega_{0}^{2}$ [Fig. \ref{fig:omega-P-E}], as well as $r(t)$ and $r^{\prime}(t)$ [Fig. \ref{fig:r-P-E}]. 
The choice of $\tau$ was such that $\omega(t)^{2}/\omega_{0}^{2}\geq0$ and $\omega^{\prime}(t)^{2}/\omega_{0}^{2}\geq0$ over the entire interval $0\leq t\leq\tau$, preventing the harmonic potential to become transiently an expulsive parabolic potential \cite{Chen-PRL-2010,Beau-Entropy-2020}. 
From Figs. \ref{fig:rho-P-E} and \ref{fig:omega-P-E}, one notes that despite the parameters $\rho(t)$ and $\rho^{\prime}(t)$, as well as $\omega(t)$ and $\omega^{\prime}(t)$, differ in the interval $0\leq t\leq\tau$, they become equal for $t>\tau$, which leads to squeezing equivalence between the TDHOs ($r_{f}=r^{\prime}_{f}$), as shown in Fig. \ref{fig:r-P-E}.

\begin{figure}[h!]
	\centering
	\subfigure[\label{fig:r-P-E-adiab}]{\epsfig{file=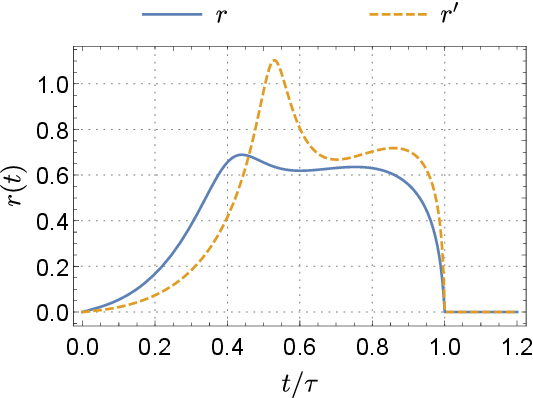,  width=0.35 \linewidth}}
	\hspace{1.0mm}
	\centering
	\subfigure[\label{fig:Q-P-E-adiab}]{\epsfig{file=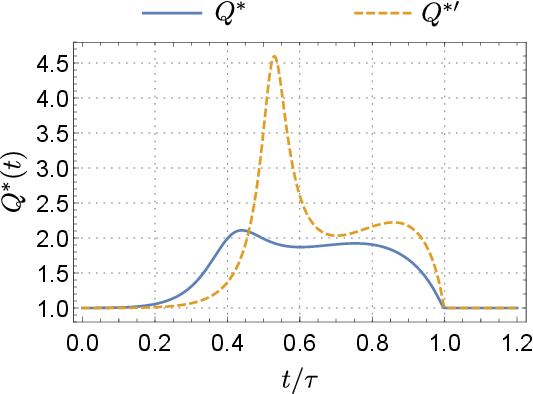, width=0.35 \linewidth}}
	\hspace{1.0mm}
	\caption{Comparison between (a) $r(t)$ and $r^{\prime}(t)$, and (b) $Q^{*}(t)$ and $Q^{*\prime}(t)$, as a function of $t/\tau$, for the case where $\rho_{\text{int}}(t)$ and $\rho_{\text{int}}^\prime(t)$ are given by Eqs. \eqref{eq:ansatz polinomial} and \eqref{eq:ansatz exponencial}, respectively. 
		The unprimed quantities are represented by the solid lines, while the primed ones by the dashed lines.
		In these figures we consider, in arbitrary units, $\omega_{0}=20$, $\omega_{f}=\omega_{f}^{\prime}=\omega_{0}/16$, $\tau=10/\omega_{0}$, $\beta=\kappa=1$, $\delta=\delta^{\prime}=1/(m_{0}\omega_{f})^{\frac{1}{2}}$, $\epsilon=\epsilon^{\prime}=0$ and $\gamma=\gamma^{\prime}=0$.}	
	\label{fig:r-Q-P-E-adiab}
\end{figure}

Different from the example discussed above, if we consider $\epsilon=\epsilon^{\prime}=0$, we obtain two adiabatic protocols.
For this situation, we show in Fig. \ref{fig:r-Q-P-E-adiab} the time evolution of $r(t)$ and $r^{\prime}(t)$ [Fig. \ref{fig:r-P-E-adiab}], and of $Q^{*}(t)$ and $Q^{*\prime}(t)$ [Fig. \ref{fig:Q-P-E-adiab}]. 
In Fig. \ref{fig:r-P-E-adiab}, one notes the occurrence of squeezing equivalence for $t>\tau$ with $r_{f}=r^{\prime}_{f}=0$, since we chose $\epsilon=\epsilon^{\prime}=0$.
Besides this, since we have adiabatic protocols in this case, when $t>\tau$ we have $Q^{*}_{f}=Q^{*\prime}_{f}=1$, as shown in Fig. \ref{fig:Q-P-E-adiab}.


\subsection{Squeezing equivalence taking into account minimization of the intermediary mean energy}\label{sec:se-energy}

In the more general context of processes in which the initial and final populations are not the same, it can be useful to analyze the minimization of the mean energy in the intermediary interval, because large transient energies can lead to anharmonic effects \cite{Chen-PRA-2010, Huang-Chaos-2020}. 
Since the quasi-optimal $\rho_{\text{int}}(t)$ that minimizes the mean energy is know (note that the conditions $\rho_{\text{int}}(t)|_{t=\tau}=\delta$ and $\dot{\rho}_{\text{int}}(t)|_{t=\tau}=\epsilon$ do not modify the parameter $\rho_{\text{int}}(t)$, found in Ref. \cite{Chen-PRA-2010}, that minimizes the mean energy), then the Ermakov-Pinney equation \eqref{eq:equação de Ermakov-Pinney} can be used to determine the frequency $\omega_{\text{int}}(t)$ that minimizes the intermediary mean energy.
The quasi-optimal $\rho_{\text{int}}(t)$ can be written as \cite{Chen-PRA-2010}
\begin{eqnarray}
    \rho_{\text{int}}(t)=\begin{cases}
	\rho_{0}\sum_{l=0}^{5}c_{l}(t/\tau)^{l}, & 0\leq t/\tau<\sigma,\\
	\rho_{0}f(t), & \sigma\leq t/\tau<1-\sigma,\\
	\rho_{0}\sum_{l=0}^{5}d_{l}(t/\tau)^{l}, & 1-\sigma\leq t/\tau\leq1,
    \end{cases}
	\label{eq:rho-min}
\end{eqnarray}
where 
\begin{eqnarray}
	f(t)=\sqrt{[B^{2}-(\omega_{0}\tau)^{2}](t/\tau)^{2}+2Bt/\tau+1},
\end{eqnarray}
in which $B=\sqrt{(\omega_{0}\tau)^{2}+\omega_{0}/\omega_{f}}-1$, $\sigma$ is related to a prescribed instant, and the coefficients $c_l$ and $d_l$ are obtained analytically from the boundary conditions \eqref{eq:cond-t0} and \eqref{eq:cond-tau} (however, the expressions are too long to be written here). 
This implies in minimization of the intermediary mean squeezing parameter [see Eq. \eqref{eq:<H>}].
Thus, imposing the squeezing equivalence conditions given in Eq. \eqref{eq:cond-equi} [or its variants \eqref{eq:equi-N}-\eqref{eq:epsilonf=epsilonf'=...epsilonf^k}], one can obtain, for instance, a function $\omega_{\text{int}}^{\prime}(t)$ so that $\omega_{\text{int}}^{\prime}(t)\neq\omega_{\text{int}}(t)$, but such that $r_{f}^{\prime}=r_{f}$.

\begin{figure}[h!]
	\centering
	\subfigure[\label{fig:rho-3}]{\epsfig{file=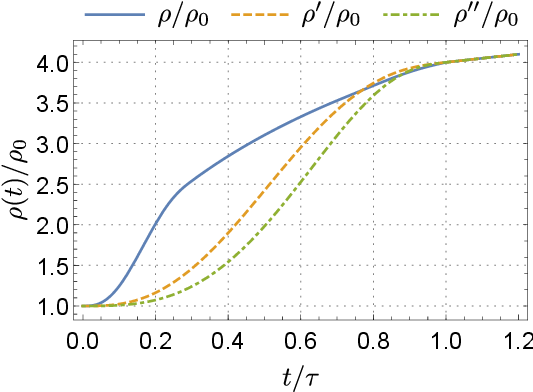,  width=0.35 \linewidth}}
	\hspace{1.0mm}
	\centering
	\subfigure[\label{fig:omega-3}]{\epsfig{file=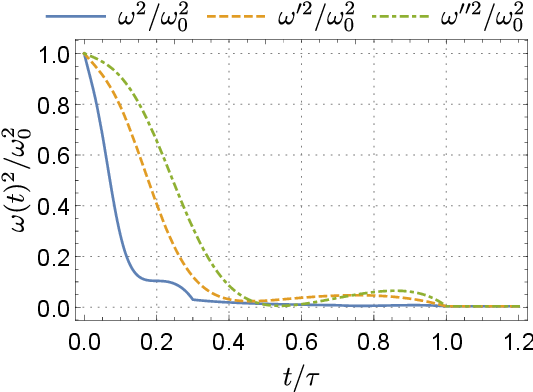, width=0.35 \linewidth}}
	\hspace{1.0mm}
	\centering
	\subfigure[\label{fig:r-3}]{\epsfig{file=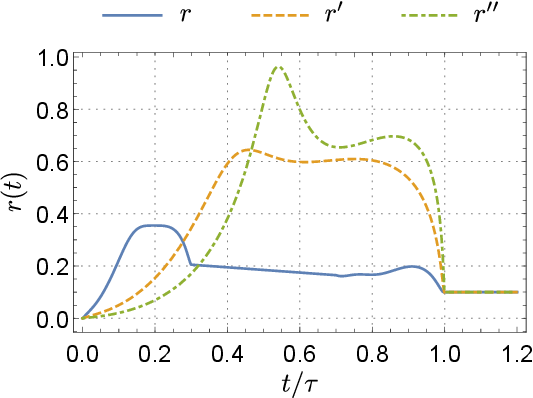, width=0.38 \linewidth}}
	\caption{Comparison between (a) $\rho(t)/\rho_{0}$, $\rho^{\prime}(t)/\rho_{0}$ and $\rho^{\prime\prime}(t)/\rho_{0}$, (b) $\omega(t)^{2}/\omega_{0}^{2}$, $\omega^{\prime}(t)^{2}/\omega_{0}^{2}$ and $\omega^{\prime\prime}(t)^{2}/\omega_{0}^{2}$, and (c) $r(t)$, $r^{\prime}(t)$ and $r^{\prime\prime}(t)$, as a function of $t/\tau$, for the case where $\rho_{\text{int}}(t)$, $\rho_{\text{int}}^{\prime}(t)$ and $\rho_{\text{int}}^{\prime\prime}(t)$ are given by Eqs. \eqref{eq:rho-min}, \eqref{eq:ansatz polinomial} and \eqref{eq:ansatz exponencial}, respectively. 
		The unprimed quantities are represented by the solid lines, while the primed ones by the dashed lines, and the double-primed ones by the dash-dotted lines.
		In these figures we consider, in arbitrary units, $\omega_{0}=20$, $\omega_{f}=\omega_{f}^{\prime}=\omega_{f}^{\prime\prime}=\omega_{0}/16$, $\sigma=0.3$, $\tau=10/\omega_{0}$, $\beta=\kappa=1$, $\delta_{k=3}=\delta_{k=1}^{\prime}=\delta_{k=1}^{\prime\prime}=1/(m_{0}\omega_{f})^{\frac{1}{2}}$, $\epsilon_{k=3}=\epsilon_{k=1}^{\prime}=\epsilon_{k=1}^{\prime\prime}=1$ and $\gamma_{k=3}=\gamma_{k=1}^{\prime}=\gamma_{k=1}^{\prime\prime}=0$.}	
	\label{fig:rho-omega-r-3}
\end{figure}

As an example of the above discussion, considering three different TDHOs, in which $\rho_{\text{int}}(t)$, $\rho_{\text{int}}^{\prime}(t)$ and $\rho_{\text{int}}^{\prime\prime}(t)$ are given by Eqs. \eqref{eq:rho-min}, \eqref{eq:ansatz polinomial} and \eqref{eq:ansatz exponencial}, respectively, we show, in Fig. \ref{fig:rho-omega-r-3}, the time evolution of $\rho(t)/\rho_{0}$, $\rho^{\prime}(t)/\rho_{0}$ and $\rho^{\prime\prime}(t)/\rho_{0}$ [Fig. \ref{fig:rho-3}], $\omega(t)^{2}/\omega_{0}^{2}$, $\omega^{\prime}(t)^{2}/\omega_{0}^{2}$ and $\omega^{\prime\prime}(t)^{2}/\omega_{0}^{2}$ [Fig. \ref{fig:omega-3}], and $r(t)$, $r^{\prime}(t)$ and $r^{\prime\prime}(t)$ [Fig. \ref{fig:r-3}] for $\epsilon_{k=3}=\epsilon_{k=1}^{\prime}=\epsilon_{k=1}^{\prime\prime}\neq0$.
Once more, despite differences between the parameters $\rho(t)$ [and $\omega(t)$], $\rho^{\prime}(t)$ [and $\omega^{\prime}(t)$] and $\rho^{\prime\prime}(t)$ [and $\omega^{\prime\prime}(t)$] within the interval $0 \leq t \leq \tau$, they converge in the interval $t > \tau$.
This results in the equality of $r(t)$, $r^{\prime}(t)$ and $r^{\prime\prime}(t)$ during the interval $t > \tau$, meaning a squeezing equivalence of the TDHOs.
As can be seen from the structure of Eq. \eqref{eq:rho-min}, this is an example of a frequency modulation involving three ($k=3$) intermediary functions [see Eqs. \eqref{eq:w-int-N} and \eqref{eq:rho-int-N}], and the squeezing equivalence was obtained with two different TDHOs involving one ($k=1$) intermediary frequency function.

\section{Final Remarks}\label{sec:final}

We pointed out that relevant problems, as the generation of squeezed states \cite{Janszky-PRA-1992} and shortcuts to adiabaticity
\cite{Chen-PRL-2010}, involve a particular case of squeezing equivalence: time-dependent quantum harmonic oscillators which have the same initial frequency $\omega_0$, the same final frequency $\omega_f$, and the same initial squeezing parameter $r_{0}=0$, subjected to different intermediary frequency modulations during a certain time interval, exhibit exactly the same final squeezing parameter $r_f=0$.
Here, we discussed a more general case of squeezing equivalence, in which a set of time-dependent quantum harmonic oscillators, subjected to different intermediary frequency modulations during a certain time interval $\tau$, can display the same non-null squeezing parameter $r_f$ after $\tau$.
Taking as basis the Lewis-Riesenfeld dynamical invariant method, we showed that the conditions for achieving this equivalence are such that, for the different oscillators, their final frequencies $\omega_{f}$, the solutions $\rho(t)$ of the Ermakov-Pinney equation \eqref{eq:equação de Ermakov-Pinney}, and $\dot{\rho}(t)$, must be equal for $t>\tau$ [Eqs. \eqref{eq:cond-equi} and \eqref{eq:wf=wf'=...wf^k}-\eqref{eq:epsilonf=epsilonf'=...epsilonf^k}].
Given that the quantum fluctuations of the position and momentum operators, quantum fluctuations of the energy value, mean number of excitations and its fluctuations, and the transition probabilities between different states, depend on the squeezing parameter [Eqs. \eqref{eq:r}-\eqref{eq:Prob-m-n-N}], the more general case of squeezing equivalence presented here implies that all these quantities will be the same for $t>\tau$, for all time-dependent quantum harmonic oscillators.

As applications, in Sec. \ref{sec:aplicações}, we demonstrated that when the interest is in controlling the forms of intermediary frequency modulations, but keeping free $r_f$ and $\tau$, this in general leads to numerical solutions to find values of $r_f$ and $\tau$, which lead to a squeezing equivalence. When applied to a particular case, this procedure recovers the squeezing equivalence found by Jansky and Adams \cite{Janszky-PRA-1992}.
On the other hand, when the interest is not in previously controlling the form of the intermediary frequencies, but rather $r_f$ and $\tau$, one can have analytical solutions to find these frequencies leading to squeezing equivalence. Particular cases of this procedure are usually applied in problems of shortcuts to adiabaticity, as done by Chen \textit{et al.} \cite{Chen-PRL-2010}, so that particular cases of the squeezing equivalence can be used as a way of obtaining different frequency modulations that result in adiabatic protocols (in this case, a null squeezing equivalence). 
In addition, we discuss the squeezing equivalence between oscillators with generic frequency modulations and one with a modulation that minimizes the intermediary mean energy. 

The more general squeezing equivalence between time-dependent quantum harmonic oscillators with different intermediary frequency modulations, discussed here, is connected to recent and important topics in the literature, which may be relevant, for example, in the area of quantum sensing, since the squeezing can be used to reduce noise in quantum systems, improving the precision of quantum measurement devices \cite{Degen-RMP-2017,Pezze-RMP-2018,Rashid-PRL-2016,Xin-PRL-2021}.


\section*{Declaration of competing interest}
The authors declare no conflict of interest.


\section*{Acknowledgements}
The authors thank Vladimir Silveira and Odemar Neto for useful discussions and careful reading of this paper.
S.S.C. and  L.Q. were supported by Coordenação de Aperfeiçoamento de Pessoal de Nível Superior - Brazil (CAPES), Finance Code 001. 
This work was partially supported by CNPq - Brazil, Processo 408735/2023-6 CNPq/MCTI.


%

\end{document}